\documentclass[preprintnumbers,amsmath,amssymbm,prd]{revtex4}
\usepackage{epsfig}
\usepackage{graphicx}

\begin{document}
\title{Bekenstein's generalized second law of thermodynamics: The
role of the hoop conjecture}
\author{Shahar Hod}
\address{The Ruppin Academic Center, Emeq Hefer 40250, Israel}
\address{ }
\address{The Hadassah Institute, Jerusalem 91010, Israel}
\date{\today}
{\it In memory of Prof. Jacob D. Bekenstein 1947---2015}

\begin{abstract}
\ \ \ Bekenstein's generalized second law (GSL) of thermodynamics
asserts that the sum of black-hole entropy,
$S_{\text{BH}}=Ac^3/4\hbar G$ (here $A$ is the black-hole surface
area), and the ordinary entropy of matter and radiation fields in
the black-hole exterior region never decreases. We here re-analyze
an intriguing gedanken experiment which was designed by Bekenstein
to challenge the GSL. In this historical gedanken experiment an
entropy-bearing box is lowered into a charged Reissner-Nordstr\"om
black hole. For the GSL to work, the resulting increase in the
black-hole surface area (entropy) must compensate for the loss of
the box's entropy. We show that {\it if} the box can be lowered
adiabatically all the way down to the black-hole horizon, as
previously assumed in the literature, then for {\it near-extremal}
black holes the resulting increase in black-hole surface-area (due
to the assimilation of the box by the black hole) may become too
small to compensate for the loss of the box's entropy. In order to
resolve this apparent violation of the GSL, we here suggest to use a
generalized version of the hoop conjecture. In particular, assuming
that a physical system of mass $M$ and electric charge $Q$ forms a
black hole if its circumference radius $r_{\text{c}}$ is equal to
(or smaller than) the corresponding Reissner-Nordstr\"om black-hole
radius $r_{\text{RN}}=M+\sqrt{M^2-Q^2}$, we prove that a new (and
larger) horizon is already formed before the entropy-bearing box
reaches the horizon of the original near-extremal black hole. This
result, which seems to have been overlooked in previous analyzes of
the composed black-hole-box system, ensures the validity of
Bekenstein's GSL in this famous gedanken experiment.
\end{abstract}
\bigskip
\maketitle

\section{Introduction}

The legend says \cite{WheLec,BekLec} that it all began with a cup of
tea and two genius physicists, Professor John Archibald Wheeler and
his young student Jacob David Bekenstein, who tried to figure out
what happens to the second law of thermodynamics when the cup goes
down a black hole.

In this gedanken experiment, the thermal entropy of the tea
disappears behind the black-hole horizon. Hence, at first glance, it
seems that the second law of thermodynamics, which states that
entropy cannot decrease, is violated in this physical process. In
particular, to external observers it seems that the entropy of the
visible universe decreases as the (entropy-bearing) object
disappears into the black hole.

It was while attempting to resolve this apparent paradox that
Bekenstein came up with the bold idea to associate entropy with
black holes -- entropy as the measure of (missing) information about
the black-hole internal state which is inaccessible to external
observers \cite{Bek73}. In particular, the formal analogy between
the second law of thermodynamics and Hawking's area theorem
\cite{Hawarea}, which states that black-hole surface area cannot
decrease \cite{Notecqa}, motivated Bekenstein to conjecture that the
required black-hole entropy \cite{Notereq} is proportional to its
surface area $A$ \cite{Bek73}:
\begin{equation}\label{Eq1}
S_{\text{BH}}={{k_{\text{B}}A}\over{4l^2_{\text{P}}}}\  .
\end{equation}
The Planck length $l_{\text{P}}=\sqrt{\hbar G/c^3}$ was introduced
into (\ref{Eq1}) by Wheeler on dimensional grounds
\cite{BekLec,Noteunit}, whereas the correct proportionality
coefficient, $1/4$, was later found by Hawking \cite{Hawco14}.

Using the conjectured proportionality (\ref{Eq1}) between black-hole
entropy and horizon area, Bekenstein proposed a generalized version
of the second law of thermodynamics \cite{Bek73}: {\it The sum of
black-hole entropy, $S_{\text{BH}}$, and the ordinary entropy of
matter and radiation fields in the black-hole exterior region, $S$,
cannot decrease}. This conjecture therefore asserts that physical
processes involving black holes are characterized by the relation
\begin{equation}\label{Eq2}
\Delta(S_{\text{BH}}+S)\geq0\  .
\end{equation}

The generalized second law of thermodynamics (GSL) provides a unique
relation between thermodynamics, gravitation, and quantum theory
\cite{Notecons}. It therefore allows us a unique glimpse into the
elusive theory of quantum gravity. It should be emphasized, however,
that despite the general agreement that the GSL reflects a
fundamental aspect of the quantum theory of gravity, there currently
exists no general proof (that is, a proof which is based on the
fundamental microscopic laws of quantum gravity) for the validity of
this principle. It is therefore of physical interest to consider
gedanken experiments in order to test the validity of the GSL in
various physical situations.

\section{Bekenstein's universal entropy bound}

In order to challenge the GSL, Bekenstein \cite{Bek73,Bek81}
analyzed a gedanken experiment in which a finite-sized object with
negligible self-gravity is assimilated into a black hole
\cite{Notecup}. In particular, Bekenstein showed that the capture of
a spherical body of proper mass $\mu$ and radius $R$ by a black hole
produces an unavoidable increase $\Delta A$ in the black-hole
surface area, whose {\it minimal} value is given by the relation
\cite{Bek73,Noteim}
\begin{equation}\label{Eq3}
(\Delta A)_{\text{min}}=8\pi\mu R\  .
\end{equation}

Taking cognizance of Eqs. (\ref{Eq1}), (\ref{Eq2}), and (\ref{Eq3}),
Bekenstein \cite{Bek73,Bek81} conjectured the existence of a
universal upper bound,
\begin{equation}\label{Eq4}
S\leq {{2\pi\mu R}\over{\hbar}}\  ,
\end{equation}
on the entropy content of physical systems with negligible
self-gravity \cite{NoteHMB,Hodb1,BekMay,Hodb2}. In particular, as
emphasized by Bekenstein \cite{Bek73,Bek81}, an entropy bound of the
form (\ref{Eq4}) ensures that the generalized second law of
thermodynamics (\ref{Eq2}) is respected in a physical process in
which a spherical body with negligible self-gravity is captured by a
black hole \cite{Notetpo}. It is worth mentioning that Bekenstein
and others \cite{Bek81,Bekev1,Bekev2,Bekev3} provided compelling
evidence that the entropy bound (\ref{Eq4}) is respected in various
physical systems in which gravity is negligible.

The main goal of the present paper is to highlight a non-trivial
aspect of Bekenstein's famous gedanken experiment \cite{Bek73}. In
particular, we shall challenge the GSL in a gedanken experiment in
which an entropy-bearing spherical body is slowly lowered into a
charged Reissner-Nordstr\"om black hole. We shall show below that
{\it if} the body can be lowered adiabatically all the way down to
the black-hole horizon, as previously assumed in the literature,
then for {\it near-extremal} black holes the unavoidable increase in
black-hole surface-area [see Eq. (\ref{Eq21}) below] may become too
small to compensate for the loss of the body's entropy
\cite{Notett}. We shall further develop a possible resolution of
this apparent paradox. In particular, we shall show that a
generalized version of the hoop conjecture \cite{Thorne} may ensure
the validity of Bekenstein's GSL in this type of gedanken
experiments.

\section{Challenging the generalized second law of thermodynamics}

We consider an entropy-bearing box of proper radius $R$ and rest
mass $\mu$ which is lowered towards a Reissner-Nordstr\"om (RN)
black hole of mass $M$ and electric charge $Q$. The external
gravitational field of the RN black-hole spacetime is described by
the line element
\begin{equation}\label{Eq5}
ds^2=-\Big(1-{{2M}\over{r}}+{{Q^2}\over{r^2}}\Big)dt^2+\Big(1-{{2M}\over{r}}+{{Q^2}\over{r^2}}\Big)^{-1}dr^2
+r^2d\Omega^2\  .
\end{equation}
The black-hole (outer and inner) horizons are located at
\begin{equation}\label{Eq6}
r_{\pm}=M\pm \sqrt{M^2-Q^2}\  .
\end{equation}
The test-particle approximation imposes the constraints
\begin{equation}\label{Eq7}
\mu\ll R\ll M\  .
\end{equation}
These strong inequalities imply that the lowered body (the
entropy-bearing box) has negligible self-gravity and that it is much
smaller than the geometric size of the black hole.

Our goal is to challenge the GSL in the most extreme situation. We
shall therefore consider the case of an entropy-bearing body which
is {\it slowly} lowered towards the black hole. As shown by
Bekenstein \cite{Bek73}, this strategy guarantees that the energy
delivered to the black hole when it swallows the body is as small as
possible \cite{Notered}. The Bekenstein strategy of lowering the
body adiabatically into the black hole also guarantees that, for
given parameters of the body, the resulting increase in the surface
area (entropy) of the black hole is minimized \cite{Bek73}.

The red-shifted energy (energy-at-infinity) of a static body which
is located at a radial coordinate $r$ in the RN black-hole spacetime
is given by \cite{Bek73}
\begin{equation}\label{Eq8}
{\cal E}(r)=\mu\sqrt{1-{{2M}\over{r}}+{{Q^2}\over{r^2}}}\ .
\end{equation}
This energy can be expressed in terms of the proper distance $l$ of
the body's center of mass above the black-hole horizon. Using the
relation \cite{Bek73}
\begin{equation}\label{Eq9}
l(r)=\int_{r_+}^{r}\sqrt{g_{rr}}dr\  ,
\end{equation}
and taking cognizance of (\ref{Eq5}), one finds the exact relation
\begin{equation}\label{Eq10}
l(r)=\sqrt{(r-r_+)(r-r_-)}+2M\ln\Big({{\sqrt{r-r_+}+\sqrt{r-r_-}}\over{\sqrt{r_+-r_-}}}\Big)\
.
\end{equation}
From (\ref{Eq10}) one finds
\begin{equation}\label{Eq11}
r(l)=r_++(r_+-r_-){{l^2}\over{4r^2_+}}[1+O(l^2/r^2_+)]\
\end{equation}
in the near-horizon $l\ll r_+$ region. Substituting (\ref{Eq11})
into (\ref{Eq8}), one finds \cite{Bek73}
\begin{equation}\label{Eq12}
{\cal E}(l)={{\mu l(r_+-r_-)}\over{2r^2_+}}\
\end{equation}
for the red-shifted energy of the box in the RN black-hole spacetime
\cite{NoteBUW,UW,Bekw,HodD}.

Suppose the entropy-bearing box is lowered slowly towards the black
hole until its center of mass lies a proper distance $l_0$ (with
$l_0\geq R$) above the horizon. The box is then released to fall
freely into the black hole. The energy (energy-at-infinity)
delivered to the black hole when it captures the body is given by
${\cal E}(l=l_0)$. The increase
\begin{equation}\label{Eq13}
\Delta M={\cal E}(l_0)={{\mu l_0(r_+-r_-)}\over{2r^2_+}}\
\end{equation}
in the mass of the RN black hole results in a change [see Eq.
(\ref{Eq6})] \cite{Notear}
\begin{equation}\label{Eq14}
\Delta A(l_0)=4\pi\Big[\Big(M+{\cal E}(l_0)+\sqrt{[M+{\cal
E}(l_0)]^2-Q^2}\Big)^2-\Big(M+\sqrt{M^2-Q^2}\Big)^2\Big]
\end{equation}
in its surface area.

From (\ref{Eq14}) one immediately realizes that $\Delta A(l_0)$ is
an increasing function of ${\cal E}(l_0)$, which also makes it an
increasing function of the dropping point $l_0$. Thus, in order to
challenge the GSL in the most extreme situation, one should release
the box to fall freely into the black hole from a point located as
close as possible to the black-hole horizon. So, we are faced with
the important question: How small can $l_0$ be made?

In his original analysis, Bekenstein \cite{Bek73} argued that the
slow descent of the body towards the black hole must stop when its
center of mass lies a proper distance $R$ \cite{Noterr,Noteqdp}
above the horizon. At this point the bottom of the box almost
touches the black-hole horizon \cite{Noteasw} and the box should
then be released to fall freely into the black hole \cite{Bek73}.
Substituting $l_0\to R$ into (\ref{Eq14}), one finds that the
minimum increase in the black-hole surface area is given by
\begin{equation}\label{Eq15}
(\Delta A)_{\text{min}}=4\pi\Big[\Big(M+{\cal E}(R)+\sqrt{[M+{\cal
E}(R)]^2-Q^2}\Big)^2-\Big(M+\sqrt{M^2-Q^2}\Big)^2\Big]\ .
\end{equation}

In the regime \cite{Noteimb}
\begin{equation}\label{Eq16}
\mu R\ll r_+(r_+-r_-)
\end{equation}
one finds [see Eq. (\ref{Eq12})] $M{\cal E}(R)\ll M^2-Q^2$, in which
case the expression (\ref{Eq15}) can be expanded in the form
\cite{Bek73}
\begin{equation}\label{Eq17}
(\Delta A)_{\text{min}}=16\pi{{r^2_+}\over{r_+-r_-}}{\cal
E}(R)\cdot\{1+O[{\cal E}(R)/(r_+-r_-)]\}\  ,
\end{equation}
which yields [see Eqs. (\ref{Eq1}) and (\ref{Eq12})] \cite{Bek73}
\begin{equation}\label{Eq18}
\Delta S_{\text{BH}}={{2\pi\mu R}\over{\hbar}}
\end{equation}
for the minimal increase in the entropy of the black hole
\cite{Noteun}. One therefore concludes \cite{Bek73} that an entropy
bound of the form (\ref{Eq4}) ensures the validity of the GSL [that
is, $\Delta S_{\text{total}}=\Delta
S_{\text{BH}}-S_{\text{body}}\geq0$] in the regime (\ref{Eq16}).

On the other hand, in the opposite regime \cite{Notedb}
\begin{equation}\label{Eq19}
r_+(r_+-r_-)\ll\mu R
\end{equation}
one finds [see Eq. (\ref{Eq12})] $M{\cal E}(R)\gg M^2-Q^2$, in which
case the expression (\ref{Eq15}) can be expanded in the form
\begin{equation}\label{Eq20}
(\Delta A)_{\text{min}}=8\sqrt{2}\pi M^{3/2}\sqrt{{\cal
E}(R)}\cdot\{1+O[\sqrt{{\cal E}(R)/M},(r_+-r_-)^2/M{\cal E}(R)]\}\
,
\end{equation}
which yields [see Eqs. (\ref{Eq1}) and (\ref{Eq12})]
\begin{equation}\label{Eq21}
\Delta
S_{\text{BH}}={{M^{3/2}\sqrt{r_+-r_-}}\over{r_+}}\cdot{{2\pi\sqrt{\mu
R}}\over{\hbar}}
\end{equation}
for the minimal increase in the entropy of the black hole. Perhaps
somewhat surprisingly, the relation (\ref{Eq21}) tells us that the
black-hole entropy increase (due to the assimilation of the body by
the black hole) can be made arbitrarily small in the extremal
$(r_+-r_-)/r_+\to 0$ limit.

In particular, one finds from (\ref{Eq21}) that, in the
near-extremal regime $(r_+-r_-)/r_+\ll1$, the increase in black-hole
entropy (due to the assimilation of the box by the black hole) may
become too small to compensate for the loss of the entropy of the
box \cite{Noteneg}. Of course, this situation is unacceptable from
the point of view of the GSL. In the next section we shall discuss a
possible resolution of this paradox which is based on (a generalized
version of) the hoop conjecture \cite{Thorne}.

\section{Bekenstein's GSL and Thorne's hoop conjecture}

We have seen that {\it if} the body can be lowered adiabatically all
the way down to the black-hole horizon, as previously assumed in the
literature, then for black holes in the {\it near-extremal} regime
(\ref{Eq19}) the resulting increase (\ref{Eq21}) in the black-hole
surface area (entropy) may become too small to compensate for the
loss of the body's entropy. In order to resolve this apparent
violation of the GSL, we shall henceforth concentrate on the
dangerous regime (\ref{Eq19}) and examine the physical consequences
of Thorne's hoop conjecture \cite{Thorne} in the context of our
gedanken experiment.

Thorne \cite{Thorne} has conjectured that a physical system of mass
$M$ forms a black hole if its circumference radius $r_{\text{c}}$ is
equal to (or smaller than) the corresponding Schwarzschild
black-hole radius $r_{\text{Sch}}=2M$ \cite{Norefr}. Interestingly,
there are several studies which support the validity of the hoop
conjecture \cite{Teuk}. It should be emphasized, however, that there
are also known counterexamples to this version of the hoop
conjecture which involve {\it charged} matter configurations
\cite{Leon,Hak}.

Hence, we would like to suggest here a natural generalization of the
hoop conjecture to the charged case: {\it A physical system of mass
$M$ and electric charge $Q$ forms a black hole if its circumference
radius $r_{\text{c}}$ is equal to (or smaller than) the
corresponding Reissner-Nordstr\"om black-hole radius
$r_{\text{RN}}=M+\sqrt{M^2-Q^2}$}. Namely, we conjecture that
\begin{equation}\label{Eq22}
r_{\text{c}}\leq M+\sqrt{M^2-Q^2}\ \  \Longrightarrow \ \
\text{Black-hole horizon exists}\  .
\end{equation}

This conjecture, if true, implies that a {\it new} horizon is formed
if the body reaches $r=r_{\text{hoop}}$, where $r_{\text{hoop}}$ is
defined by the Reissner-Nordstr\"om relation [see Eq. (\ref{Eq6})]
\begin{equation}\label{Eq23}
r_{\text{hoop}}=M+{\cal E}(r_{\text{hoop}})+\sqrt{[M+{\cal
E}(r_{\text{hoop}})]^2-Q^2}\ .
\end{equation}
Substituting (\ref{Eq8}) into (\ref{Eq23}), one finds
\begin{equation}\label{Eq24}
r_{\text{hoop}}=M+\sqrt{M^2-Q^2+4\mu^2}\
\end{equation}
for the location of the new (and larger) horizon.

Taking cognizance of (\ref{Eq7}), one realizes that the dangerous
regime (\ref{Eq19}) is characterized by the relations
\begin{equation}\label{Eq25}
r_+-r_-\ll\mu\ll r_+\  ,
\end{equation}
Thus, the radius (\ref{Eq24}) of the new horizon can be written in
the form
\begin{equation}\label{Eq26}
r_{\text{hoop}}=M+2\mu\{1+O[(r_+-r_-)^2/\mu^2]\}\  .
\end{equation}
Substituting (\ref{Eq26}) into (\ref{Eq10}), one finds
\begin{equation}\label{Eq27}
l(r_{\text{hoop}})=M\cdot\Big[\ln\Big({{\mu}\over{r_+-r_-}}\Big)+O(1)\Big]\gg
M\gg R\ .
\end{equation}
From the inequality $l(r_{\text{hoop}})\gg R$ [see (\ref{Eq27})] one
learns that, in the dangerous regime (\ref{Eq19}), the new horizon
\cite{Notenh} is already formed {\it before} the body reaches the
horizon of the original near-extremal black hole.

The formation of the new (and larger) horizon (\ref{Eq24}) implies
an increase [see Eqs. (\ref{Eq6}) and (\ref{Eq26})]
\begin{equation}\label{Eq28}
\Delta A=4\pi(r^2_{\text{hoop}}-r^2_+)=16\pi
M\mu\{1+O[\mu/r_+,(r_+-r_-)^2/\mu^2]\}\
\end{equation}
in black-hole surface area. The corresponding increase in black-hole
entropy is given by \cite{Noteqw}
\begin{equation}\label{Eq29}
\Delta S_{\text{BH}}={{4\pi M\mu}\over{\hbar}}\gg {{2\pi\mu
R}\over{\hbar}}\  .
\end{equation}
From (\ref{Eq29}) one learns that the (generalized) hoop conjecture
(\ref{Eq22}) \cite{Thorne}, together with Bekenstein's universal
entropy bound (\ref{Eq4}), ensure the validity of the GSL [that is,
$\Delta S_{\text{total}}=\Delta S_{\text{BH}}-S_{\text{body}}\geq0$]
in the dangerous regime (\ref{Eq19}).

\section{Summary}

Historically, the idea to associate the black-hole surface area with
entropy [see (\ref{Eq1})] was suggested by Bekenstein \cite{Bek73}
in order to save the validity of the second law of thermodynamics in
a gedanken experiment in which an entropy-bearing object falls into
a black hole \cite{WheLec,BekLec}.

Following this bold conjecture, Bekenstein \cite{Bek73} proposed a
generalized version of the second law of thermodynamics, according
to which the sum of black-hole entropy (given by $A/4\hbar$) and the
ordinary entropy of matter and radiation fields in the black-hole
exterior region never decreases [see (\ref{Eq2})]. In particular, it
was shown by Bekenstein \cite{Bek73} that, in the regime
$r_+(r_+-r_-)\gg\mu R$ of black holes which are far from extremality
\cite{Noteim,Noteimb}, an entropy bound of the form (\ref{Eq4})
ensures the validity of the GSL in the famous gedanken experiment in
which an entropy-bearing spherical body of radius $R$ and proper
mass $\mu$ is slowly lowered into a black hole \cite{Noteen}.

In the present paper we have highlighted a non-trivial aspect of
Bekenstein's historical gedanken experiment: We have shown that {\it
if} the body can be lowered slowly all the way down to the
black-hole horizon, as previously assumed in the literature, then
for {\it near-extremal} black holes the resulting increase in
black-hole surface-area (due to the assimilation of the body by the
black hole) may become too small to compensate for the loss of the
body's entropy [see Eq. (\ref{Eq21})].

In order to resolve this apparent violation of the GSL, we have
suggested to use a (generalized) version of the hoop conjecture
\cite{Thorne}. In particular, assuming that a physical system of
mass $M$ and electric charge $Q$ forms a black hole if its
circumference radius $r_{\text{c}}$ is equal to (or smaller than)
the corresponding Reissner-Nordstr\"om black-hole radius
$r_{\text{RN}}=M+\sqrt{M^2-Q^2}$ [see Eq. (\ref{Eq22})], we have
proved that a new (and {\it larger}) horizon is already formed {\it
before} the body reaches the horizon of the original near-extremal
black hole.

Finally, we have explicitly shown that the increase (\ref{Eq29}) in
black-hole entropy, which is a direct consequence of the
(generalized) hoop conjecture (\ref{Eq22}), when combined with the
Bekenstein entropy bound (\ref{Eq4}), ensures the validity of
Bekenstein's GSL in this historical gedanken experiment
\cite{Notetdb}.

\bigskip
\noindent
{\bf ACKNOWLEDGMENTS}
\bigskip

In memory of Prof. Jacob D. Bekenstein (1947--2015), whose seminal
papers have inspired my research.
\newline
This research is supported by the Carmel Science Foundation. I would
like to thank Yael Oren, Arbel M. Ongo, Ayelet B. Lata, and Alona B.
Tea for helpful discussions.


\begin{thebibliography}{99}

\bibitem{WheLec} See ``Interview with John Wheeler" at
https://www.youtube.com/watch?v=C0EsJPpX5lc.

\bibitem{BekLec} See J. D. Bekenstein at
https://www.youtube.com/watch?v=XkLrmRVmGZ4 (In Hebrew).

\bibitem{Bek73} J. D. Bekenstein, Lett. Nuov. Cim. {\bf 4}, 737 (1972); J. D.
Bekenstein, Phys. Rev. D {\bf 7}, 2333 (1973); J. D. Bekenstein,
Phys. Rev. D {\bf 9}, 3292 (1974).

\bibitem{Hawarea} S. W. Hawking, Phys. Rev. Lett. {\bf 26}, 1344
(1971). See also: D. Christodoulou, Phys. Rev. Lett. {\bf 25}, 1596
(1970); D. Christodoulou and R. Ruffini, Phys. Rev. D {\bf 4}, 3552
(1971).

\bibitem{Notecqa} It is important to emphasize that Hawking's area
theorem is based on the {\it classical} weak (positive) energy
condition \cite{Hawarea}. Thus, quantum processes which violate the
positive energy condition can reduce the surface area of a black
hole.

\bibitem{Notereq} As discussed above, this black-hole entropy is
required in order to save the validity of the second law of
thermodynamics in Bekenstein's gedanken experiment
\cite{WheLec,BekLec}.

\bibitem{Noteunit} We shall henceforth use natural units in which
$G=c=k_{\text{B}}=1$.

\bibitem{Hawco14} S. W. Hawking, Commun. Math. Phys. {\bf 43}, 199
(1975).

\bibitem{Notecons} Note that the expression (\ref{Eq1}) for the
black-hole entropy contains all known fundamental constants of
nature: the Newton gravitational constant $G$, the speed of light
$c$, and the quantum Planck constant $\hbar$ (in addition, this
relation also contains the thermodynamic Boltzmann constant
$k_{\text{B}}$).

\bibitem{Bek81} J. D. Bekenstein, Phys. Rev. D {\bf 23}, 287 (1981).

\bibitem{Notecup} Remember the cup of tea whose disappearance (along with its
entropy) into the black hole initiated the research on black-hole
thermodynamics \cite{WheLec,BekLec}.

\bibitem{Noteim} It is worth emphasizing that in deriving the lower bound (\ref{Eq3}),
it was implicitly assumed in \cite{Bek73} that the black hole is far
from extremality in the sense that $r_+(r_+-r_-)\gg\mu R$, where
$r_{\pm}$ are the black-hole horizon radii [see Eq. (\ref{Eq16})
below]. In the present paper we shall rigorously analyze this
near-extremal case. In particular, we shall show below that
near-extremal black holes in the regime $r_+(r_+-r_-)\ll\mu R$ pose
the most dangerous threat to the GSL.

\bibitem{NoteHMB} It is worth mentioning that it was later shown in \cite{Hodb1,BekMay,Hodb2} that
the upper bound (\ref{Eq4}) can be improved for rotating and charged
physical systems.

\bibitem{Hodb1} S. Hod, Phys. Rev. D {\bf 61}, 024018 (2000) [arXiv:gr-qc/9901035].

\bibitem{BekMay} J. D. Bekenstein and A. E. Mayo, Phys. Rev. D {\bf 61}, 024022 (2000).

\bibitem{Hodb2} S. Hod, Phys. Rev. D {\bf 61}, 024023 (2000) [arXiv:gr-qc/9903011].

\bibitem{Notetpo} That is, the unavoidable increase (\ref{Eq3}) in the
black-hole surface-area (entropy) guarantees that the GSL
(\ref{Eq2}) is respected in this type of gedanken experiments
provided the entropy $S$ of the body is bounded from above by
(\ref{Eq4}).

\bibitem{Bekev1} J. D. Bekenstein, Phys. Rev. D {\bf 30}, 1669 (1984);
J. D. Bekenstein, Phys. Rev. D {\bf 49}, 1912 (1994).

\bibitem{Bekev2} J. D. Bekenstein and E. I. Guendelman, Phys. Rev. D {\bf 35}, 716
(1987).

\bibitem{Bekev3} J. D. Bekenstein and M. Schiffer, Int. J. Mod. Phys. C {\bf 1}, 355
(1990); M. Schiffer and J. D. Bekenstein, Phys. Rev. D {\bf 39},
1109 (1989); C. Eling and J. D. Bekenstein, Phys. Rev. D {\bf 79},
024019 (2009).

\bibitem{Notett} To the best of our knowledge, this fact [see Eq. (\ref{Eq21}) below]
has been overlooked in previous analyzes of this famous gedanken
experiment.

\bibitem{Thorne} K. S. Thorne, in {\it Magic without Magic: John Archibald Wheeler}, edited by J. Klauder (Freeman,
San Francisco, 1972).

\bibitem{Notered} Note that the mass-energy of the box (energy as measured by asymptotic observers)
is {\it red-shifted} during its slow descent towards the black hole
[see Eq. (\ref{Eq8}) below].

\bibitem{NoteBUW} It is worth noting that, due to quantum buoyancy effects
\cite{UW}, there is an additional quantum contribution to the energy
of the (slowly lowered) body. However, as shown by Bekenstein
\cite{Bekw} (see also \cite{HodD}), for macroscopic and mesoscopic
objects in the regime $\eta\equiv R/R_{\text{C}}\gg1$ (here
$R_{\text{C}}\equiv\hbar/\mu$ is the Compton length of the body),
the quantum contribution to the energy of the body is very small.
Specifically, for objects in the regime $R\gg R_{\text{C}}$ (note
that this inequality is already valid for atomic-sized systems
\cite{Bekw}), quantum buoyancy effects are suppressed by a large
factor of order $O(\eta^{1/3})$ relative to the mass-energy
(\ref{Eq12}) of the body \cite{Bekw}.

\bibitem{UW} W. G. Unruh and R. M. Wald, Phys. Rev. D {\bf 25}, 942
(1982).

\bibitem{Bekw} J. D. Bekenstein, Phys. Rev. D {\bf 49}, 1912 (1994);
J. D. Bekenstein, Phys. Rev. D {\bf 60}, 124010 (1999).

\bibitem{HodD} S. Hod, Jour. of High Energy Phys. {\bf 1012}, 033
(2010) [arXiv:1101.3151].

\bibitem{Notear} Here we have used the expression $A=4\pi r^2_+$ for
the surface area of the RN black hole.

\bibitem{Noterr} Recall that $R$ is the proper radius of the body.
As emphasized by Bekenstein \cite{Bek73}, the expression
(\ref{Eq12}) for the energy of the body in the black-hole spacetime
is only valid when every part of it is still {\it outside} the
black-hole horizon.

\bibitem{Noteqdp} It is worth noting that quantum buoyancy effects
\cite{UW} shift the optimal dropping point of the body (that is, the
radial location for which the energy of the body is minimized) to a
point slightly above $l=R$. Specifically, as shown in \cite{Bekw}
(see also \cite{HodD}), for objects in the regime $\eta\equiv
R/R_{\text{C}}\gg1$ (as emphasized earlier, this inequality is
already valid for atomic-sized systems \cite{Bekw}), the optimal
dropping point of the body lies a proper distance
$l_0=R[1+O(\eta^{-1/3})]\simeq R$ above the black-hole horizon.

\bibitem{Noteasw} That is, at this point the bottom of the box begins to be swallowed by the
black hole.

\bibitem{Noteimb} It is worth emphasizing that the strong inequality
(\ref{Eq16}) was implicitly assumed in the original analysis of
Bekenstein \cite{Bek73}. In particular, the {\it first-order}
relation $\Delta A={{16\pi r^2_+}\over{r_+-r_-}}\Delta M$ used in
\cite{Bek73} [see, in particular, Eqs. (8) and (A15) of J. D.
Bekenstein, Phys. Rev. D {\bf 7}, 2333 (1973)] is only valid in the
$M\Delta M\ll (r_+-r_-)^2$ regime [or equivalently, in the regime
$\mu R\ll r_+(r_+-r_-)$, see Eq. (\ref{Eq13})].

\bibitem{Noteun} As emphasized by Bekenstein \cite{Bek73}, the black-hole entropy
increase (\ref{Eq18}) in the regime (\ref{Eq16}) \cite{Noteimb} is
universal in the sense that it is independent of the black-hole
parameters.

\bibitem{Notedb} It is worth emphasizing again \cite{Noteim,Noteimb}
that this regime was not analyzed in Bekenstein's seminal paper
\cite{Bek73}.

\bibitem{Noteneg} Namely, the relation (\ref{Eq21}) suggests $\Delta
S_{\text{total}}=\Delta S_{\text{BH}}-S_{\text{body}}\to
-S_{\text{body}}<0$ in the near-extremal $(r_+-r_-)/r_+\to 0$ limit.

\bibitem{Norefr} More precisely, for non-spherically symmetric quasi-static systems,
like the one we consider here, the hoop conjecture \cite{Thorne}
requires the circumference radius of the system to be equal to (or
smaller than) the corresponding Schwarzschild black-hole radius in
{\it every} direction.

\bibitem{Teuk}  See A. M. Abrahams, K. R. Heiderich, S. L. Shapiro and S. A. Teukolsky,
Phys. Rev. D {\bf 46}, 2452 (1992) and references therein.

\bibitem{Leon} See J. P. de Le\'on, Gen. Relativ. and Grav. {\bf 19},
289 (1987) and references therein.

\bibitem{Hak} H. Andreasson, Commun. Math. Phys. {\bf 288}, 715 (2009).

\bibitem{Notenh} It is worth emphasizing again that
this new horizon is expected to be formed according to the
(generalized) hoop conjecture \cite{Thorne}.

\bibitem{Noteqw} The last inequality in (\ref{Eq29}) follows from
(\ref{Eq7}).

\bibitem{Noteen} That is, the entropy of the box disappears from the visible universe,
but the increase (\ref{Eq18}) in black-hole entropy guarantees that
the GSL is respected: $\Delta S_{\text{total}}=\Delta
S_{\text{BH}}-S_{\text{body}}\geq0$.

\bibitem{Notetdb} To the best of our knowledge, the important role
played by the (generalized) hoop conjecture in ensuring the validity
of the GSL has not been discussed in the literature so far.

\end{thebibliography}
\end{document}